\documentclass[aip,jcp,amsmath,amssymb,preprint]{revtex4-1}

\usepackage{graphicx}
\usepackage{dcolumn}
\usepackage{bm}
\usepackage{xcolor}
\usepackage{amsfonts,amsmath,amsthm,amssymb,mathtools}

\newcommand*{\p}{\partial}

\newcommand*{\nablab}{\bar{\nabla}}

\newcommand*{\di}{\iint d\textbf{r}_1\,d\textbf{r}_2\;}

\begin{document}
\preprint{AIP/JCP/}

\title[]{Calculation of the molecular integrals with the range-separated correlation factor}

\author{Micha\l~Silkowski}
\author{Micha\l~Lesiuk}
\email{lesiuk@tiger.chem.uw.edu.pl.}
\author{Robert Moszynski}
\affiliation{Faculty of Chemistry, University of Warsaw, Pasteura 1, 02-093 Warsaw, Poland}

\date{\today}

\begin{abstract}
Explicitly correlated quantum chemical calculations require calculations of five types of two-electron 
integrals beyond the standard electron repulsion integrals. We present a novel scheme, which utilises general ideas 
of the McMurchie-Davidson technique, to compute these integrals when the so-called ``range-separated'' correlation 
factor is used. This correlation factor combines the well-known short range behaviour resulting from the electronic 
cusp condition, with the exact long-range asymptotics derived for the helium atom [M. Lesiuk, B. Jeziorski, and R. 
Moszynski, J. Chem. Phys. \textbf{139}, 134102 (2013)]. Almost all steps of the presented 
procedure are formulated recursively, so that an efficient implementation and control of the precision are possible. 
Additionally, the present formulation is very flexible and general, and it allows for use of an arbitrary correlation 
factor in the electronic structure calculations with minor or no changes.
\end{abstract}

\maketitle

\section{Introduction}
\label{sec:intro}

Most of the traditional \emph{ab initio} quantum chemistry methods represent the electronic wavefunction as a linear 
combination of the Slater determinants. This leads to severe difficulties in describing the Coulomb hole because the
electronic cusp condition requires that the electronic wavefunction behaves linearly in the interelectronic distance,
$r_{12}$, near the coalescence points of the electrons \cite{kato57,pack66,furnais05,hoffmann92,king96}. To overcome
this
difficulty explicitly correlated methods have been introduced which include the $r_{12}$ factor in the 
wavefunction\cite{hattig12,kong12,tenno12}. Typically, the aforementioned $r_{12}$ dependence is implemented trough the
product of two occupied molecular orbitals with the correlation factor, $f_{12}$. The correlation factor depends solely
on $r_{12}$ and makes it possible to satisfy the cusp condition, so that the description of the Coulomb hole is
dramatically improved. However, an important issue in the explicitly correlated methods is the choice of the correlation
factor itself. Originally, the straightforward choice $f_{12}=r_{12}$ was made which gave rise to the class of R12
methods \cite{kutz85,klopper87,kutz91,noga92,noga94,klopper03,szalewicz10,szalewicz82,szalewicz99,szalewicz03}.
Later, Manby generalised the theory and allowed for an arbitrary correlation factor to be used\cite{may04}.
It was also found that the choice $f_{12}=r_{12}$ is responsible for the bulk of the error in the early R12 methods \cite{may05}. In 2004,
Ten-no proposed the exponential correlation factor \cite{tenno04b}, which is nowadays routinely applied in various
explicitly correlated approaches \cite{shiozaki11a,shiozaki11b,torheyden09,tenno07,shiozaki10,kedzuch11,martinez10}, and
was shown to improve the results significantly as compared to R12. This suggests that the
actual form of the correlation factor is important and one should continue searching for a more optimal form of
$f_{12}$. In fact, Tew and Klopper \cite{tew05} investigated numerically the shape of the correlation factor for the
helium atom and helium-like ions and compared it with several simple analytic forms. They found that in the short-range
$r_{12}$ regime the exponential correlation factor of Ten-no is close to the optimal but they did not have access to
moderate and large $r_{12}$ distances. In a recent work, the form of correlation factor for the helium atom
has been reconsidered by using purely analytic methods \cite{lesiuk2013}. It has rigorously been found that the
asymptotic (long-range) form of the correlation factor is $f_{12} \rightarrow r_{12}^\rho e^{B r_{12}}$, as $r_{12}
\rightarrow \infty$, for the helium atom with several plausible models of the electronic wavefunction. It is obvious
that none of the correlation factors which are used presently obeys this asymptotic behaviour. Therefore, we proposed the
so-called \emph{range-separated} correlation factor\cite{lesiuk2013} where the short- and long-range $r_{12}$ regimes are approximated by
different formulae and sewed together by using a switching function. A detailed functional form will be specified
further in the paper.

One may wonder about the motivation behind the introduction of the range-separated correlation factor.
From a brief inspection of the problem it appears that the long-range part of the correlation factor is completely 
irrelevant and the only thing that matters is the ability to model the correlation hole. However, recent results have 
proven that standard basis sets used in the conventional (determinants-based) calculations are suboptimal for
the F12 methods \cite{peterson08,peterson10a,peterson10b}, where the form of the correlation factor is arbitrary. This
is due to the fact that the cusp region is described
reasonably well by the explicitly correlated part of the trial wavefunction and basis set functions are mainly necessary
to model the longer-range part of the electronic wavefunction. Therefore, one can expect that a correlation
factor which includes a physically motivated long-range ingredient can further reduce the basis set requirements
of the current F12 approaches. However, the largest gain in the accuracy can be reached with methods where the
correlation factor is responsible for a dominant fraction of the so-called dynamical correlation effects. A prominent
method of this type is the geminal-augmented CASSCF method of Martinez \emph{et al.} (G-CASSCF) \cite{martinez10}. In
this method the wavefunction is modelled as a standard CASSCF expansion multiplied by a proper explicitly correlated
part. Therefore, the situation is clear: the better and more physically motivated the correlation factor is, the larger
percentage of the (dynamical) correlation energy is retrieved which directly improves the results. Note additionally
that ``guide'' wavefunctions used in the Quantum Monte Carlo (QMC) methods \cite{luchow00,foulkes01,austin12} also take
the form of the Slater determinants multiplied by the explicitly correlated Jastrow factor. Unfortunately, it is not
clear whether the present form of $f(r_{12})$ introduced in the context of F12 methods is advantageous when applied to
QMC as well. Nonetheless, this possibility is worth investigating, particularly in the light of recent remarkable
improvements in the QMC technology and quality of the results that can routinely be obtained (see Refs.
\onlinecite{luchow05,luchow08,dubecky13a,todd08,galeka07,benedek06,korth11,dubecky10,dubecky14,dubecky13b} for
representative
examples).

However, before the actual performance of the range-separated correlation factor can be tested, one needs to evaluate 
the necessary matrix elements. Efficient and stable evaluation of the two-electron integrals 
is a constant challenge for the electronic structure theory and a lot of efforts have been devoted to the development 
of fast and reliable integral codes. For the standard electron repulsion integrals including the Coulomb potential, 
$1/r_{12}$, several procedures were developed in the works of Dupuis, Rys, and King 
\cite{dupuis76a,dupuis76b,dupuis83}, McMurchie and Davidson \cite{mcmurchie78,helgaker92}, Pople and Hehre 
\cite{pople78}, Obara and Saika \cite{obara86}, and others \cite{schlegel82,headgordon88,lindh91,hamilton91,johnson93}. 
The review paper of Gill \cite{gill94} is very informative in this respect. Introduction of the explicitly correlated 
methods in quantum chemistry raised the number of types of two-electron integrals which need to be 
routinely evaluated. Typically, each specific form of the correlation factor was handled separately by modifying the 
existing algorithms. The works of Klopper and R\"{o}hse \cite{klopper96}, Bearpart \emph{et al.} \cite{bearpart91}, 
Valeev and Schaefer \cite{valeev00}, Ten-no \cite{tenno04b,tenno07}, Samson \emph{et al.} \cite{samson02}, and Weber 
and Daul \cite{weber04} are prominent examples of this approach. In this paper we consider the evaluation of all 
two-electron integrals that are necessary to perform F12 calculations with the range-separated correlation factor
\cite{lesiuk2013}. 
We show that these integrals can be evaluated by applying the general idea of the McMurchie-Davidson recursive
scheme \cite{mcmurchie78,helgaker92}. Our work provides a large level of generality. In fact, with minor or 
no changes the present scheme can be applied to an \emph{arbitrary} correlation factor. Ahlrichs \cite{ahlrichs06} 
presented a method to compute two-electron integrals with any kernel within the Obara-Saika technique, but his general 
recursive procedure is multidimensional and complicated, especially when the analytical form of the correlation factor 
is complicated. As far as we know, the present level of generality has never been achieved thus far within the 
McMurchie-Davidson scheme.

This paper is organised as follows. In Sec. \ref{subsec:cart} we introduce the two-electron integrals 
necessary in the F12 calculations and perform their initial reduction. Next, in Sec. \ref{subsec:cart} the 
McMurchie-Davidson technique is applied to express these integrals trough the derivatives of the so-called 
basic integrals with respect to the orbitals locations. In Sec. \ref{subsec:basic} we establish an analytical equation 
for the basic integrals. Sections \ref{subsec:rr} and \ref{subsec:radial} detail the evaluation of the required 
derivatives of the basic integrals with the help of the Hobson theorem. Section \ref{sec:rs12} presents the application 
of the developed scheme to the range-separated correlation factor. The generating integrals $S(\alpha,\beta,\gamma)$ 
are introduced and their evaluation is discussed. Finally, Sec. \ref{sec:conclusion} concludes our paper.

\section{Integrals with a general correlation factor}
\label{sec:secgen}

\subsection{Cartesian two-electron integrals}
\label{subsec:cart}

It is now a well established fact that in the F12 theories the following six types of two-electron integrals 
appear:
\begin{align}
\label{eq:I1}
I_1 &=(ab \vert r^{-1}_{12} \vert cd) = \di
\Phi_a^\ast(\textbf{r}_1)\,
\Phi_c^\ast(\textbf{r}_2)\,
r_{12}^{-1}\,
\Phi_b(\textbf{r}_1)\,
\Phi_d(\textbf{r}_2),\\
\label{eq:I2}
I_2 &=(ab \vert r^{-1}_{12}f_{12} \vert cd) = \di
\Phi_a^\ast(\textbf{r}_1)\,
\Phi_c^\ast(\textbf{r}_2)\,
r_{12}^{-1}f_{12}(r_{12})\,
\Phi_b(\textbf{r}_1)\,
\Phi_d(\textbf{r}_2),\\
\label{eq:I3}
I_3 &=(ab \vert f_{12} \vert cd) = \di
\Phi_a^\ast(\textbf{r}_1)\,
\Phi_c^\ast(\textbf{r}_2)\,
f_{12}(r_{12})\,
\Phi_b(\textbf{r}_1)\,
\Phi_d(\textbf{r}_2), \\
\label{eq:I4}
I_4 &=(ab \vert f_{12}f'_{12} \vert cd) = \di 
\Phi_a^\ast(\textbf{r}_1)\,
\Phi_c^\ast(\textbf{r}_2)\,
f_{12}(r_{12})f'_{12}(r_{12})\,
\Phi_b(\textbf{r}_1)\,
\Phi_d(\textbf{r}_2),\\
\label{eq:I5}
I_5 &=(ab \vert \lbrack f_{12}{,}\hat{T_1} \rbrack \vert cd) = \di 
\Phi_a^\ast(\textbf{r}_1)\,
\Phi_c^\ast(\textbf{r}_2)\,
\lbrack f_{12}(r_{12}){,}\hat{T_1} \rbrack \,
\Phi_b(\textbf{r}_1)\,
\Phi_d(\textbf{r}_2),\\
\label{eq:I6}
I_6 &=(ab \vert \lbrack \lbrack  f_{12}{,}\hat{T_1} \rbrack{,} f'_{12} \rbrack \vert cd) = \di 
\Phi_a^\ast(\textbf{r}_1)\,
\Phi_c^\ast(\textbf{r}_2)\,
\lbrack \lbrack f_{12}(r_{12}){,}\hat{T_1} \rbrack{,}f'_{12}(r_{12})\rbrack \,
\Phi_b(\textbf{r}_1)\,
\Phi_d(\textbf{r}_2),
\end{align}
where $\hat{T}_1=-\frac{1}{2}\nabla_1^2$ stands for the kinetic energy operator of the first 
electron. We have additionally increased the flexibility of our approach by allowing two different correlation 
factors, $f_{12}$ and $f'_{12}$, to appear simultaneously. Functions on the brackets are the (unnormalised) primitive
Cartesian Gaussian-type orbitals (GTOs)
\begin{align}
\label{eq:basis}
\Phi(\textbf{r},a,\textbf{A})=x^i_{A}y^k_{A}z^m_{A} \exp\left(-ar_{A}^2\right),
\end{align}
where $a$ denotes the exponent and $\textbf{A}=(A_x,A_y,A_z)$ is the vector specifying the location of the orbital. In 
the above expression $x_A = x - A_x$, and similarly for the other components, and $r_A^2 = x_A^2 + y_A^2 + z_A^2$. 
The same notation is preserved for the remaining orbitals in Eqs. (\ref{eq:I1})-(\ref{eq:I6}), which 
are located at the centres $B$, $C$, $D$, and have the respective exponents $b$, $c$, $d$. 

In all explicitly correlated approaches three- and possibly four-electron integrals appear, as the 
result of the action of the Coulomb, exchange, or strong orthogonality
operators on the pair functions \cite{kutz85,klopper87,kutz91,noga92,noga94,klopper03,szalewicz10,szalewicz03}. 
However, all these difficult
integrals are usually not evaluated explicitly but rather approximated \emph{e.g.} by the insertion of the resolution 
of identity (RI) in terms of an auxiliary basis set \cite{klopper02,tenno03,valeev04a}. This approximation leaves only 
two-electron integrals in the form (\ref{eq:I1})-(\ref{eq:I6}). Other techniques for handling many-electron integrals 
were also suggested \cite{may04,persson96,persson97,dahle01,tenno04a,saito01,komornicki11}. Let us also note that in
some explicitly correlated approaches two- and three-electron nuclear attraction integrals are required. This includes,
\emph{e.g.}, the G-CASSCF method of Martinez \emph{et al.} \cite{martinez10}. The same types of integrals are necessary
for the explicit electron-nucleus correlation treatments of Hammes-Schiffer \emph{et al.} (see Ref.
[\onlinecite{hammes08}] and references therein). In this work we concentrate on the integrals
(\ref{eq:I1})-(\ref{eq:I6}) which are necessary for the standard explicitly correlated MP2 or CC methods. Nonetheless,
we believe that extensions to other special classes of the integrals are feasible and will be considered in further
works.

Let us first simplify the $r_{12}$-dependent kernels in Eqs. (\ref{eq:I5}) and (\ref{eq:I6}). Firstly, by applying the 
second Green's identity in Eq. (\ref{eq:I5}), the integral $I_5$ can be rewritten as
\begin{align}
\label{eq:I5_}
 I_5 = \frac{1}{2} \left(\nabla_{\textbf{A}}^2-\nabla_{\textbf{B}}^2 \right) I_3.
\end{align}
Secondly, the kernel of Eq. (\ref{eq:I6}) can be simplified by noting that
\begin{align}
\lbrack \lbrack f_{12}(r_{12}){,}\hat{T_1} \rbrack{,}f'_{12}(r_{12})\rbrack = (\nabla_1
f_{12})\cdot(\nabla_1 f'_{12}) = \frac{\p f_{12} }{\p r_{12}}\frac{\p f'_{12}}{\p r_{12}},
\end{align}
so that the difficult double commutator was replaced by the product of the derivatives of the correlation factors.
Thus, Eq. (\ref{eq:I6}) takes now a more explicit form
\begin{align}
\label{eq:I6_}
I_6 =(ab \vert \frac{\p f_{12} }{\p r_{12}}\frac{\p f'_{12}}{\p r_{12}} \vert cd).
\end{align}
The above rearrangements are valid for any correlation factors, $f_{12}$ and $f'_{12}$, for which the 
initial integrals exist.

\subsection{Application of the McMurchie-Davidson scheme}
\label{subsec:murchie}

Once the simplifications described above have been introduced, we can attack the integrals 
(\ref{eq:I1})-(\ref{eq:I6}) by using the McMurchie-Davidson (MD) scheme. Note that our presentation of this
well-established technique is necessarily brief and we consequently do not enter into the issues already 
solved in the literature. For a more elaborate discussion of the MD method we address the reader to the original papers
\cite{mcmurchie78,helgaker92}, or the review by Gill \cite{gill94}.

The main idea behind the MD method is the following expansion of the product of two Cartesian Gaussian orbitals
\begin{align}
\label{eq:AB}
\begin{split}
 \Phi(\textbf{r}_1,a,\textbf{A})~\Phi(\textbf{r}_1,b,\textbf{B})&=\sum_{t=0}^{i+j} E_t^{ij} \sum_{u=0}^{k+l} 
E_u^{kl}\sum_{v=0}^{m+n} E_v^{mn} \Lambda_{tuv}(\textbf{r}_1,p,\textbf{P})=\\
&=\sum_{tuv} E_{tuv}^{ab} 
\Lambda_{tuv}(\textbf{r}_1,p,\textbf{P}),
\end{split}
\end{align}
where the latter formulae is simply a shorthand notation for the former. In the above expression the so-called Hermite 
Gaussian functions were introduced
\begin{align}
\label{eq:Lambda}
\Lambda_{tuv}(\textbf{r}_1,p,\textbf{P})=\frac{d^t}{dP_x^t}\frac{d^u}{dP_y^u}\frac{d^v}{dP_z^v}\exp\left(-pr_{1P}
^2\right),
\end{align}
with the new exponent, $p=a+b$, and the displaced centre, $\textbf{P}=\frac{a\textbf{A}+b\textbf{B}}{p}$. 
The coefficients $E_t^{ij}$ obey the following recursion relations which can be used to calculate them efficiently
\begin{align}
 &E_0^{00}=\exp\left(-\frac{ab}{a+b}(A_x-B_x)^2\right),
 \label{eq:Erec0}
 \\ &E_0^{i+1,j}= -\frac{b}{p}(A_x-B_x)E_0^{ij}+E_1^{ij},
 \label{eq:Ereci}
 \\ &E_0^{i,j+1}= \frac{a}{p}(A_x-B_x)E_0^{ij}+E_1^{ij}E_0,
 \label{eq:Erecj}
 \\ &E_t^{ij}= \frac{1}{2pt}\left(iE_{t-1}^{i-1,j}+jE_{t-1}^{i,j-1}\right).
  \label{eq:Erect}
\end{align}
Of course, a very similar formulae hold for the product of orbitals of the second electron
with $p$ replaced by $q=c+d$, and $\textbf{P}$ by $\textbf{Q}=\frac{c\textbf{C}+d\textbf{D}}{q}$.
Let us now insert the expression (\ref{eq:AB}) into Eqs. (\ref{eq:I1})-(\ref{eq:I4}) and 
(\ref{eq:I6}), taking into account the simplified formula (\ref{eq:I6_}). The integral $I_5$ will be treated 
separately. Note that the differentiation with respect to the coordinates of $\textbf{P}$ can be replaced by 
the differentiation with respect to the components of $-\textbf{Q}$, simply because of the translational
invariance of the resulting expressions. After some rearrangements we arrive at the following formulae:
\begin{align}
\label{eq:ii}
 I_i = \sum_{tuv} E_{tuv}^{ab} (-1)^{t+u+v} \sum_{t'u'v'} E_{t'u'v'}^{cd} R_i^{t+t',u+u',v+v'},\;\;\;i=1,2,3,4,6,
\end{align}
where
\begin{align}
\label{eq:ri}
 R_i^{tuv} = \frac{d^t}{dQ_x^t}\frac{d^u}{dQ_y^u}\frac{d^v}{dQ_z^v}B_i,
\end{align}
and $B_i$ are the basic integrals that are given by the following generic expression
\begin{align}
\label{eq:bi}
B_i = \di e^{-pr_{1P}^2}k_i(r_{12})e^{-qr_{2Q}^2}.
\end{align}
In the above equation $k_i(r_{12})$ are the kernel functions dependent solely on $r_{12}$ and, in general, different 
for each $i=1,2,3,4,6$. In the most general case, the kernel functions take the following form
\begin{align}
\label{eq:k1}
k_1(r_{12}) &= r^{-1}_{12},\\
\label{eq:k2}
k_2(r_{12}) &= r_{12}^{-1}f_{12}(r_{12}),\\
\label{eq:k3}
k_3(r_{12}) &= f_{12}(r_{12}),\\
\label{eq:k4}
k_4(r_{12}) &= f_{12}(r_{12})f'_{12}(r_{12}),\\
\label{eq:k6}
k_6(r_{12}) &= \frac{\p f_{12} }{\p r_{12}}\frac{\p f'_{12}}{\p r_{12}}.
\end{align}

As mentioned above, the integral $I_5$ does not follow the scheme given above and requires a separate method. Let 
us recall Eq. (\ref{eq:I5_}) and follow the approach advocated by Helgaker \emph{et al.} \cite{helgaker92} in the
context of the geometrical derivatives. Firstly, the differentiation with respect to the coordinates of $\textbf{A}$
and $\textbf{B}$ is rewritten as
\begin{align}
 \frac{d}{dA_x}=-\frac{a}{p}\frac{d}{dQ_x}+\frac{d}{d{AB}_x},\\
 \frac{d}{dB_x}=-\frac{b}{p}\frac{d}{dQ_x}-\frac{d}{d{AB}_x},
\end{align}
where the vector $\textbf{AB}$ is simply $\textbf{A}-\textbf{B}$. With the above equations at hand, it becomes simple 
to express the operator $\frac{1}{2} \left(\nabla_{\textbf{A}}^2-\nabla_{\textbf{B}}^2 \right)$ present in Eq. 
(\ref{eq:I5_}) trough the 
derivatives with respect to the components of $\textbf{Q}$ and $\textbf{AB}$. Next, by applying it to the integral
$I_3$, given by Eq. (\ref{eq:ii}) with $i=3$, one obtains
\begin{align}
\label{eq:I5byI3_}
\begin{split}
 I_5&=\frac{a-b}{2p} \sum_{tuv} E_{tuv}^{ab} (-1)^{t+u+v} \sum_{t'u'v'} E_{t'u'v'}^{cd} \nabla_{\textbf{Q}}^2 
R_3^{t+t',u+u',v+v'} \\
 &- \frac{a-b}{p} \sum_{tuv} \vec{\nabla}_{AB} E_{tuv}^{ab} (-1)^{t+u+v} \cdot \sum_{t'u'v'} E_{t'u'v'}^{cd} 
\vec{\nabla}_{\textbf{Q}} R_3^{t+t',u+u',v+v'},
\end{split}
\end{align}
where $\vec{\nabla}_{\textbf{X}}$ and $\nabla_{\textbf{X}}^2$ are the gradient and Laplace operators with respect to 
$\textbf{X}$, $\textbf{X}\in\{ \textbf{Q}, \textbf{AB} \}$.
The main difficulty now lies in the evaluation of the derivatives of $R_3^{tuv}$ and the 
coefficients $E_{tuv}$. In the case of $R_3^{tuv}$ one simply recalls Eq. (\ref{eq:ri}) which leads to
\begin{align}
&\nabla_{\textbf{Q}}^2 R_3^{tuv} = R_3^{t+2,uv} + R_3^{t,u+2,v} + R_3^{tu,v+2},\\
&\vec{\nabla}_{\textbf{Q}} R_3^{tuv} = \left[ R_3^{t+1,uv}, R_3^{t,u+1,v}, R_3^{tu,v+1} \right].
\end{align}
Additionally, as pointed out by Samson \emph{et al.} \cite{samson02}, the action of the $\vec{\nabla}_{\textbf{AB}}$ on 
$E_{tuv}^{ab}$ can be resolved by the following recurrence relations:
\begin{align}
 &F_0^{00} =-\frac{2ab}{p} AB_x E_0^{00},\\
 &F_0^{i+1,j} =-\frac{b}{p} AB_x F_0^{ij}+F_1^{ij}-\frac{b}{p} E_0^{ij}, \\
 &F_0^{i,j+1} =\frac{a}{p} AB_x F_0^{ij}+F_1^{ij}+\frac{a}{p} E_0^{ij}, \\
 &F_t^{ij} = \frac{1}{2pt}\left( iF_{t-1}^{i-1,j}+jF_{t-1}^{i,j-1} \right),
\end{align}
where $F_t^{ij}=\frac{dE_t^{ij}}{dAB_x}$ and a similar notation ($F_u^{kl}$ and $F_v^{mn}$) holds for the components $y$ 
and $z$. The above expressions are straightforwardly obtained by a direct differentiation of Eqs. 
\eqref{eq:Erec0}-\eqref{eq:Erect}. To sum up, the integral $I_5$ is also expressed as a linear combination of the
quantities $R_3^{tuv}$.

Let us summarise the progress made in this section by using the MD scheme. Firstly, the integrals $I_1$-$I_4$ 
and $I_6$ have been reduced to linear combinations of the $R_i^{tuv}$ functions. We have shown that the latter
quantities are the derivatives of $B_i$ with respect to the components of $\textbf{Q}$. Finally,
the integral $I_5$, resistant to the standard approach, has also been written down as a linear combination of 
$R_i^{tuv}$. All the coefficients appearing in
the expressions obey simple recursive schemes and thus can efficiently be computed. Therefore, the remaining 
issues which need to be addressed are the evaluation of the basic integrals, $B_i$, and their differentiation with
respect to the components of $\textbf{Q}$. Note that the approach adopted in this section is similar, but more 
general, to the work of Samson \emph{et al.} \cite{samson02} However, at this point our derivations
separate completely.

\subsection{Basic integrals}
\label{subsec:basic}

In order to evaluate the quantities $R_i^{tuv}$, $i=1,2,3,4,6$, one has to establish an analytical 
expression for the basic integrals, $B_i$, given by Eq. (\ref{eq:bi}). The main difficulty lies in the fact that the 
form of the correlation factors, and thus the form of the kernels $k_i$, has not been specified yet. In typical 
situations, an analogue of the basic integral is evaluated with use of the integral transforms of the kernel 
functions. For instance, in the case when $k_1=r_{12}^{-1}$ the following transformation can be adopted
\begin{align}
r_{12}^{-1}=\frac{2}{\sqrt{\pi}}\int_0^{\infty} \exp(-t^2r_{12}^2) dt,  
\end{align}
and a similar one was used by Ten-no for the exponential correlation factor \cite{tenno04b,tenno07}. In the case of an
arbitrary correlation
factor such an integral transform may not exist. Thus, a different approach
needs to be devised. As already noted before, all basic integrals are translationally invariant. In other words, these
integrals are dependent only on the length of the vector $\textbf{P}-\textbf{Q}$ but not on its individual components. 
Therefore, without any loss of generality, let us make a shift and put consequently $\textbf{P}=0$ further in the 
paper. The basic integrals now take the following form
\begin{align}
\label{eq:bi_}
B_i = \di e^{-pr_1^2}k_i(r_{12})e^{-qr_{2Q}^2},
\end{align}
where the kernel functions, $k_i$, are defined exactly in the same way as in the previous subsection. Let us now recall 
the formula for translation of the $1s$ Gaussian function
\begin{align}
\label{eq:gaussshift}
e^{-qr_{2Q}^2}=4\pi e^{-qQ^2}e^{-qr_2^2}\sum_{l=0}^{\infty}i_l(2qQr_2)\sum_{m=-l}^{+l}Y_{lm}^{\ast}(\hat{Q})Y_{lm}(\hat{
r_2}),
\end{align}
where $Q=\vert\textbf{Q}\vert$, $i_l(z)=\sqrt{\frac{\pi}{2z}}I_{n+\frac{1}{2}}(z)$, $I_\nu(z)$ are the modified 
Bessel functions of the first kind \cite{stegun72}, and $Y_{lm}(\hat{r})$ are the spherical harmonics\cite{stegun72}. In 
fact, this formula is very well known to the quantum chemists community, as it is routinely used in the
evaluation of the matrix elements of the effective core potentials (see, for example, the work of Kahn \emph{et al.}
\cite{kahn76}). Let us now insert Eq. (\ref{eq:gaussshift}) into Eq. 
(\ref{eq:bi_}). An important observation here is that all terms, apart from $l=0$ and $m=0$, vanish due to the 
spherical symmetry of the integrand. Therefore, one is left with
\begin{align}
\label{eq:bis}
B_i = e^{-qQ^2} \di e^{-pr_1^2-qr_2^2}\,\frac{\sinh(2qQr_2)}{2qQr_2}\,k_i(r_{12}),
\end{align}
where we have additionally used the well-known formula $i_0(z)=\sinh(z)/z$. Note that the integrand in Eq. 
(\ref{eq:bis}) is essentially a one-centre distribution. Therefore, it is convenient to utilise the variables $r_1$, 
$r_2$ and $r_{12}$ supplemented with three Euler angles. After the change of variables and straightforward 
integration over the angles the following triple integral remains
\begin{align}
\label{eq:bis_}
B_i = 8\pi^2\,\frac{e^{-qQ^2}}{2qQ}\int_0^{\infty}dr_{12}\,k_i(r_{12})\,r_{12}\int_0^{\infty}dr_2\,e^{-qr_2^2}
\sinh(2qQr_2)\int_{|r_2-r_{12}|}^{r_2+r_{12}}dr_1\,r_1\,e^{-pr_1^2}.
\end{align}
The inner two integrations in the above expression are elementary. After some rearrangements one arrives at
\begin{align}
\label{eq:bfinal}
B_i = \sqrt{\frac{\pi^5}{p+q}}\frac{1}{qp} \int_0^{\infty} dr_{12}\,k_i(r_{12})\,r_{12} \left[
\frac{e^{-\frac{pq}{p+q}(r_{12}-Q)^2}-e^{-\frac{pq}{p+q}(r_{12}+Q)^2}}{Q}\right].
\end{align}
The above expression is universal \emph{i.e.} it is valid for any kernel for which the initial integrals exist. 
Additionally, only the terms in the square brackets in Eq. (\ref{eq:bfinal}) are dependent on $Q$.
This fact is crucial for developments of the further sections.

\subsection{Evaluation of the integrals $R_i^{tuv}$}
\label{subsec:rr}
The integrals $R_i^{tuv}$, defined formally as derivatives of the basic integrals, $B_i$, with respect to the 
components of $\textbf{Q}$ can now formally be calculated by a consecutive differentiation of Eq. (\ref{eq:bfinal})
with respect to $Q_x$, $Q_y$ and $Q_z$. However, this straightforward approach is quite cumbersome since 
the differentiation produces very lengthly expressions. Apart from that, the integrals $B_i$ depend only on the length 
of $\textbf{Q}$, but not on its individual components, which automatically suggests that considerable simplifications 
are possible.

Let us designate by $g(Q)$ an arbitrary well-behaved and sufficiently many times differentiable function which is 
dependent solely on the length of the vector $\textbf{Q}$.
Now, the spherical tensor gradient operators, $\mathcal{Y}_{lm}(\nablab)$, are obtained 
by taking the explicit formulae for the regular solid spherical harmonics, $r^l Y_{lm}(\hat{r})$, and replacing the 
Cartesian 
coordinates, $x$, $y$, $z$, by the differentials, $\frac{\partial}{\partial x}$, $\frac{\partial}{\partial y}$, 
$\frac{\partial}{\partial z}$, respectively. A special case of the Hobson theorem \cite{hobson65,hobson92} reads
\begin{align}
\label{eq:Ylm}
\mathcal{Y}_{lm}(\nablab_Q) g(Q) = Q^l \left[ \left(\frac{1}{Q}\frac{d}{dQ}\right)^l g(Q) \right] Y_{lm}(\hat{Q}).
\end{align}
Exactly the same formula is valid for the real versions of the spherical harmonics. One simply takes a 
combination of the above expression with $m$ and $-m$. This simple observation removes the need to use 
complex variables in actual computations. Concerning Eq. (\ref{eq:Ylm}), see the works of Weniger and Steinborn 
\cite{weniger83,weniger85}, for a more elaborate discussion of the Hobson theorem.

The Hobson theorem makes it simple to operate with $\mathcal{Y}_{lm}(\nablab)$, but the original definition of 
$R_i^{tuv}$ is given in terms of the derivatives with respect to Cartesian coordinates. Therefore, one has to relate an 
arbitrary differential 
with respect to the Cartesian coordinates with the spherical tensor gradient operators. Obviously, the transformation
formulae are exactly the same as between Cartesian coordinates and the 
spherical harmonics, which follows directly from the definition of $\mathcal{Y}_{lm}(\nablab)$. In our notation, the
necessary transformation reads
\begin{align}
\begin{split}
\label{eq:QxQyQz}
\frac{d^t}{dQ_x^t}\frac{d^u}{dQ_y^u}\frac{d^v}{dQ_z^v}&= 
N(l,t,u,v) \sum_{l\leq t+u+v} c(l,m,t,u,v)\, \mathcal{Y}_{lm}(\nablab).
\end{split}
\end{align}
in accordance with the work of Schlegel \cite{schlegel95}. For completeness, the analytical expressions for the 
coefficients $N(l,t,u,v)$ and $c(l,m,t,u,v)$ are given in Supplemental Material \cite{supplement}. The former are fairly
simple and can be computed on the fly. The latter are very sparse \emph{i.e.} a huge number of combinations of the
parameters gives zero value. Therefore, they can be calculated in advance and stored in the memory for further use.

Making use of Eqs. (\ref{eq:Ylm}) and (\ref{eq:QxQyQz}) one can write
\begin{align}
\begin{split}
R_i^{tuv} &= \frac{d^t}{dQ_x^t}\frac{d^u}{dQ_y^u}\frac{d^v}{dQ_z^v} R_i^{000} =
N(l,t,u,v) \sum_{l\leq t+u+v} c(l,m,t,u,v)\, \mathcal{Y}_{lm}(\nablab) R_i^{000} =\\
&= N(l,t,u,v) \sum_{l\leq t+u+v} c(l,m,t,u,v)\, Q^l \left[ \left(\frac{1}{Q}\frac{d}{dQ}\right)^l B_i \right] 
Y_{lm}(\hat{Q}),
\end{split}
\end{align}
where the identity $R_i^{000}=B_i$ has been used. By inspection of the above expression one immediately 
recognises that only the derivatives of $B_i$ with respect to $Q$ are necessary. They are considered in the next 
subsection. Other quantities appearing in the above expression are the spherical harmonics, but they can 
efficiently be computed with help of the well-known recursive formulae \cite{stegun72}.

\subsection{Evaluation of the radial derivatives}
\label{subsec:radial}

Due to the advances presented in the previous subsection it becomes clear that a major effort of the whole scheme 
lies in evaluation of the following terms
\begin{align}
\label{eq:DQ}
 Q^l \mathcal{D}_Q^l B_i,
\end{align}
where the notation $\mathcal{D}_Q=\frac{1}{Q}\frac{d}{dQ}$ has been implied. In the evaluation of the 
differentials (\ref{eq:DQ}) an additional attention must be paid to the numerical issues. In fact, the $Q$-dependent 
factor in Eq. (\ref{eq:bfinal}) consists of two terms which are of opposite signs and nearly equal in magnitude when the 
value of $Q$ is small. Therefore, a significant digital erosion can be expected when Eq. (\ref{eq:bfinal}) is evaluated
as it stands for small $Q$. This problem is also largely magnified when a direct differentiation of Eq.
(\ref{eq:bfinal}) is attempted to calculate the radial derivatives, Eq. (\ref{eq:DQ}). Therefore, we need to adopt a 
special approach when the value of $Q$ is small.

\subsubsection{The case of small $Q$}

Let us rewrite the $Q$-dependent part of the integrand in Eq. (\ref{eq:bfinal}) as follows
\begin{align}
\label{eq:qterm}
\frac{e^{-\xi (r_{12}-Q)^2}-e^{-\xi (r_{12}+Q)^2}}{Q} = 
\frac{e^{-\xi(r_{12}^2+Q^2)}}{Q}\sinh(2\xi r_{12}Q) 
= \frac{e^{-\xi(r_{12}^2+Q^2)}}{Q} \sum_{n=0}^\infty \frac{\left(2\xi r_{12}Q\right)^{2n+1}}{(2n+1)!},
\end{align}
where $\xi=pq/(p+q)$.
The above infinite series is everywhere convergent. Therefore, it can be inserted back into Eq.
(\ref{eq:bfinal}) and the order of summation and integration can be interchanged. This leads to the following series 
expansion of 
$B_i$ around $Q=0$
\begin{align}
\label{eq:biqsmall}
B_i = 2\pi\,e^{-\xi Q^2}\left(\frac{\pi}{p+q}\right)^{3/2} \sum_{n=0}^\infty \frac{\left(2\xi 
Q\right)^{2n}}{(2n+1)!} \int_0^{\infty} dr_{12}\,k_i(r_{12})\,r_{12}^{2n+2}\,e^{-\xi r_{12}^2}.
\end{align}

Passing to the evaluation of the differentials (\ref{eq:DQ}), it is now straightforward to apply the operator $Q^l 
\mathcal{D}_Q^l$ to both sides of Eq. (\ref{eq:biqsmall}) which gives
\begin{align}
\label{eq:bidersmall}
Q^l \mathcal{D}_Q^l B_i = 
2\pi Q^l \left(\frac{\pi}{p+q}\right)^{3/2} \sum_{n=0}^\infty \frac{\left(2\xi 
\right)^{2n}}{(2n+1)!}\,\mathcal{D}_Q^l\left(Q^{2n}e^{-\xi Q^2}\right) \int_0^{\infty} 
dr_{12}\,k_i(r_{12})\,r_{12}^{2n+2}\,e^{-\xi r_{12}^2}.
\end{align}
The action of $\mathcal{D}_Q^l$ on the term $Q^{2n}e^{-\xi Q^2}$ is conveniently resolved by means of the following 
recursion relation
\begin{align}
\mathcal{D}_Q^l \left(Q^{2n}e^{-\xi Q^2}\right) = n\,\mathcal{D}_Q^{l-1} \left(Q^{2n-2}e^{-\xi Q^2}\right)
-2\xi\,\mathcal{D}_Q^{l-1} \left(Q^{2n}e^{-\xi Q^2}\right)
\end{align}
which is sufficiently numerically stable for small $Q$. Let us conclude by noting that the series expansion 
(\ref{eq:bidersmall}) converges reasonably well provided that the values of $Q$ and $\xi$ are simultaneously small. In 
the opposite situation the approach developed in the next subsection is appropriate.

\subsubsection{The case of moderate or large $Q$}

In this subsection we consider the evaluation of the radial derivatives, $Q^l \mathcal{D}_Q^l B_i$, by means of the 
recursive techniques. Since the problem of small values of $Q$ has already been solved, we can perform separation of 
the two terms in the square brackets in Eq. (\ref{eq:bfinal}), so that
\begin{align}
\label{eq:hjkl}
\prescript{i}{}{\mathcal{H}_{jkl}^\pm} = \sqrt{\frac{\pi^5}{p+q}}\frac{1}{qp} \int_0^\infty 
dr_{12}\,r_{12}^j\,k_i(r_{12})\,
\mathcal{D}_Q^l\left(\frac{e^{-\xi(r_{12}\pm Q)^2}}{Q^k}\right),
\end{align}
and 
\begin{align}
\label{eq:recbiq}
Q^l \mathcal{D}_Q^l B_i = Q^l \left(\prescript{i}{}{\mathcal{H}_{11l}^-} - \prescript{i}{}{\mathcal{H}_{11l}^+}\right).
\end{align}
We assume here that the values of $\prescript{i}{}{\mathcal{H}_{j00}^\pm}$ are available at the start and their 
evaluation will be considered in details later. It is obvious that $\prescript{i}{}{\mathcal{H}_{jk0}^\pm}$ are 
trivial to obtain from the initial $\prescript{i}{}{\mathcal{H}_{j00}^\pm}$, so our task is to increase the value 
of $l$ at the cost of $j$ and $k$. Let us first inspect the action of $\mathcal{D}_Q$ on the $Q$-dependent part of the 
integrand in Eq. (\ref{eq:hjkl}) with $l=0$
\begin{align}
 \mathcal{D}_Q\left(\frac{e^{-\xi(r_{12}\pm Q)^2}}{Q^k}\right)=
 \mp\,2\xi r_{12}\frac{e^{-\xi(r_{12}\pm Q)^2}}{Q^{k+1}}-
 2\xi \frac{e^{-\xi(r_{12}\pm Q)^2}}{Q^k}-
 k\frac{e^{-\xi(r_{12}\pm Q)^2}}{Q^{k+2}}.
\end{align}
Let us now multiply both sides of the above expression by $r_{12}^j\,k_i(r_{12})$ and integrate over $r_{12}$ on the 
interval $[0,+\infty]$. One obtains the following relation
\begin{align}
\prescript{i}{}{\mathcal{H}_{jk1}^\pm}=
\mp\,2\xi\, \prescript{i}{}{\mathcal{H}_{j+1,k+1,0}^\pm}
-2\xi\, \prescript{i}{}{\mathcal{H}_{jk0}^\pm}
-k\, \prescript{i}{}{\mathcal{H}_{j,k+2,0}^\pm},
\end{align}
which can further be generalised by applying $\mathcal{D}_Q^{l-1}$ to both sides
\begin{align}
\label{eq:finalrec}
\prescript{i}{}{\mathcal{H}_{jkl}^\pm}=
\mp\,2\xi\, \prescript{i}{}{\mathcal{H}_{j+1,k+1,l-1}^\pm}
-2\xi\, \prescript{i}{}{\mathcal{H}_{jk,l-1}^\pm}
-k\, \prescript{i}{}{\mathcal{H}_{j,k+2,l-1}^\pm}.
\end{align}
This gives us the desired recursion relation which can be used to calculate 
$\prescript{i}{}{\mathcal{H}_{11l}^\pm}$ starting only with $\prescript{i}{}{\mathcal{H}_{jk0}^\pm}$. Note that the 
recursive process (\ref{eq:finalrec}) is rather simple, despite operating in three dimensions, and can efficiently be
implemented. Therefore, the presented recursive method offers an attractive way for the evaluation of the radial 
derivatives.


\section{The range-separated correlation factor}
\label{sec:rs12}
\subsection{Preliminaries}
\label{subsec:pre}
Let us now consider the so-called range-separated correlation factor \cite{lesiuk2013}
\begin{align}
\label{eq:rs12}
f_{12}(r_{12};c_0,\rho,B,\mu) = \left(1+\frac{1}{2}r_{12} \right) e^{-\mu r_{12}^2}+
S_n\left(\mu r_{12}^2\right) c_0r_{12}^\rho e^{B r_{12}},
\end{align}
where $c_0,\rho,B,\mu>0$ are variational parameters and
\begin{align}
\label{eq:ttdamp}
S_n(x) =  1-e^{-x}\sum_{k=0}^n \frac{x^k}{k!},
\end{align}
is the Tang-Toennies damping function \cite{tang84}. The integer $n$ is assumed to be greater than $-\frac{\rho}{2}-1$, 
if $\rho<-2$, or $n=0$ otherwise [note that $S_n(x)=\mathcal{O}(x^{n+1})$]. It is clear that the term 
$\left(1+\frac{1}{2}r_{12} \right) 
e^{-\mu r_{12}^2}$ is responsible for the short-range correlation and necessary to fulfil the cusp condition. Even 
greater flexibility can be achieved if the exponential formula of Ten-no is used in the short-range 
regime\cite{lesiuk2013}. The long-range part $r_{12}^\rho e^{B r_{12}}$ represents the correct asymptotics, as found 
for the helium atom\cite{lesiuk2013}. The role of the Gaussian and 
Tang-Toennies damping functions is to interpolate smoothly between these two regimes, providing at the same time a 
large degree of flexibility. It is important to note that the parameters $\rho$, $B$ can be arbitrary real numbers. 
In particular, it was found for the helium atom that the value of $B$ is always positive and the value of $\rho$ 
depends significantly on the orbital part of the wave function. In the case when $\rho$ is negative the damping 
function $S_n$ removes the corresponding singularity as $r_{12}\rightarrow 0$. This is the reason why the aforementioned
constraint on $n$ strictly holds. The choice of the parameter $\mu$ is particularly important as it controls the
transition between the short- and long-range regimes. The optimal value of $\mu$ will be determined by benchmarking on a
set of model systems, similarly as it was done for the parameter $\gamma$ present in the exponential correlation factor
of Ten-no\cite{tenno03,tenno04a,tenno04b,tew05}.

Before the theory developed in the previous section can be applied, one mathematical difficulty needs to be addressed. 
Let us consider the kernel $k_3(r_{12})$ given by Eq. (\ref{eq:k3}). To calculate the necessary integrals by 
using the recursive method, one starts with the quantities $\prescript{3}{}{\mathcal{H}_{j00}^\pm}$. One can formally 
insert the explicit form of the correlation factor, Eq. (\ref{eq:rs12}), into the expression (\ref{eq:hjkl}). 
After several simplifications, this leads to the formula
\begin{align}
\label{eq:hj00}
\begin{split}
\prescript{3}{}{\mathcal{H}_{j00}^\pm} &= \mathcal{N}
\Bigg[
\int_0^\infty dr_{12}\,r_{12}^j\, e^{-(\mu+\xi) r_{12}^2}\,e^{\mp 2\xi r_{12}Q}
+\frac{1}{2}\int_0^\infty dr_{12}\,r_{12}^{j+1}\,e^{-(\mu+\xi) r_{12}^2} \,e^{\mp 2\xi r_{12}Q}\\
&+c_0\int_0^\infty dr_{12}\,r_{12}^{j+\rho}\, S_n\left(\mu r_{12}^2\right) e^{\left(B \mp 2\xi Q\right)r_{12}}
\Bigg],
\end{split}
\end{align}
where $\mathcal{N}=\sqrt{\frac{\pi^5}{p+q}}\frac{e^{-\xi Q^2}}{qp}$. An obvious step now is to insert the explicit form 
of the damping function, Eq. (\ref{eq:ttdamp}), into the above expression. However, the summation present in Eq. 
(\ref{eq:ttdamp}) cannot be straightforwardly interchanged with the integration in Eq. (\ref{eq:hj00}) since some of 
the resulting integrals can become singular. Therefore, we need to change the integration range from $[0,+\infty]$ to 
$[\varepsilon,+\infty]$ and take the limit $\varepsilon\rightarrow 0^+$ once the separation of the individual terms is 
performed. This gives
\begin{align}
\label{eq:hj00eps}
\begin{split}
&\prescript{3}{}{\mathcal{H}_{j00}^\pm} = \mathcal{N}
\Bigg[
\int_0^\infty dr_{12}\,r_{12}^j\, e^{-(\mu+\xi) r_{12}^2}\,e^{\mp 2\xi r_{12}Q}
+\frac{1}{2}\int_0^\infty dr_{12}\,r_{12}^{j+1}\,e^{-(\mu+\xi) r_{12}^2} \,e^{\mp 2\xi r_{12}Q} \\
&+c_0\lim_{\varepsilon\rightarrow 0^+}\Bigg(
\int_{\varepsilon}^\infty dr_{12}\,r_{12}^{j+\rho}\, e^{\left(B \mp 2\xi Q\right)r_{12}}
-\sum_{k=0}^n \frac{\mu^k}{k!}
\int_{\varepsilon}^\infty dr_{12}\,r_{12}^{j+\rho+2k}\, e^{\left(B \mp 2\xi Q\right)r_{12}} e^{-\mu r_{12}^2}\Bigg)
\Bigg].
\end{split}
\end{align}
These findings suggest that the all necessary integrals can be expressed trough the following family of functions
\begin{align}
\label{eq:sila}
S(\alpha,\beta,\gamma)=\mathfrak{R}\int_0^{\infty} dx\, x^{\alpha}\exp(\beta x-\gamma x^2),
\end{align}
where $\alpha$, $\beta$ are arbitrary real numbers and $\gamma \geq 0$ and $\mathfrak{R}$ denotes the finite (regular)
part of the integral. The precise meaning of the symbol $\mathfrak{R}$ is defined as follows. When $\alpha>-1$ the
integrals (\ref{eq:sila}) are regular and can be treated with standard methods. For $\alpha \leq -1$ they
are divergent and have to be understood differently. Therefore, when $\alpha \leq -1$ we proceed as follows. First, the
integration range is 
changed to $[\varepsilon,+\infty]$ and this modified integral is evaluated. Next, the small $\varepsilon$ expansion is 
considered and one obtains an analytical result which consist of a regular, $\varepsilon$-independent part 
of the integral plus terms which contain $\varepsilon$. The latter terms split into two groups: 
vanishing as the limit $\varepsilon\rightarrow0$ is taken and singular terms which are \emph{finite} combinations of
inverse 
powers and possibly logarithms of $\varepsilon$. In the present regularisation scheme all singular 
terms are simply dropped and the limit $\varepsilon\rightarrow0$ is taken. The result is a finite, regular part of the 
otherwise diverging integrals. This approach is fully justified because in all expressions for the relevant 
integrals, such as Eq. (\ref{eq:hj00eps}), the integrals $S$ appear in specific combinations, inherited after the form 
of the damping functions. In these combinations, all singular terms must cancel out since the expressions for the 
initial matrix elements are regular. In other words, there is no need to keep track of the terms containing 
$\varepsilon$ because we know in advance that they must cancel out exactly when the final expression, such as 
(\ref{eq:hj00eps}), is evaluated.

As a simple example, let us consider $S(\alpha,\beta,\gamma)$ with $\alpha=-2$, $\gamma=0$ and $\beta<0$. Integration 
over the range $[\varepsilon,+\infty]$, followed by a small $\varepsilon$ expansion gives
\begin{align}
\int_{\varepsilon}^{\infty} dx\, \frac{e^{\beta x}}{x^2} = 
-\beta \bigg[ \gamma_E + \log(-\beta) - 1 \bigg] + \frac{1}{\varepsilon} - \beta \log(\varepsilon) - 
\frac{\beta^2}{2}\varepsilon + \mathcal{O}(\varepsilon^2),
\end{align}
where $\gamma_E$ is the Euler-Mascheroni constant. One can see that the exemplary integral consists of the regular part 
(terms in the square bracket), a finite number of singular terms ($\frac{1}{\varepsilon}$ and $\log \varepsilon$) and 
an infinite number of terms which vanish as $\varepsilon\rightarrow0$. As the singular terms are dropped and the limit 
$\varepsilon\rightarrow0$ is taken one obtains in our notation
\begin{align}
S(-2,\beta,0) = \mathfrak{R} \int_{0}^{\infty} dx\, \frac{e^{\beta x}}{x^2} = -\beta \bigg[ \gamma_E + \log(-\beta) - 1 
\bigg].
\end{align}
Clearly, this technique is a fairly straightforward way to manage the spurious singularities. 

Let us note in passing that in practical situations the form of the correlation factor is fixed \emph{i.e.} one 
adopts some recommended values of the parameters in Eq. (\ref{eq:rs12}). By recalling Eq. (\ref{eq:hj00eps}) as an 
example, one finds that in actual computations a set of basic integrals of the form 
$S(\alpha_0+N,\beta,\gamma)$ is required, where $\alpha_0$ is a real parameter, $0<\alpha_0<1$, and $N$ is an arbitrary 
integer. The values of $\beta$ and $\gamma$ depend on the nonlinear parameters of the GTOs, whilst $\alpha_0$ is 
inherited over the adopted formula for $f_{12}$. This observation is rather obvious, but important in the context of 
the next subsections where calculation of $S(\alpha,\beta,\gamma)$ is discussed in details.

In the Supplemental Material \cite{supplement} we give the analytical forms of the kernel functions, $k_i$, with the
range-separated correlation factor, and express the corresponding basic integrals, $B_i$, through
$S(\alpha,\beta,\gamma)$. These expressions are long but very regular, and can be turned into an efficient computer
code.

\subsection{Calculation of $S(\alpha,\beta,\gamma)$ for $\alpha > -1$}
\label{subsec:silaint}

In the case when the value of $\alpha$ is larger than $-1$ the integrals $S(\alpha,\beta,\gamma)$ can be reduced to the 
Tricomi's confluent hypergeometric function, $\mathcal{U}(a,b,z)$, by a simple variable transformation 
$t=\sqrt{2\gamma}x$. In this way one obtains
\begin{align}
\label{eq:suabz}
S(\alpha,\beta,\gamma)=(2\sqrt{\gamma})^{-\alpha-1}\Gamma(\alpha+1)\,
\mathcal{U}(\frac{\alpha+1}{2},\frac{1}{2},\frac{\beta^2} {4\gamma}).
\end{align}
At this point the problem is solved because methods of calculation of $\mathcal{U}(a,b,z)$ for arbitrary complex
values of the parameters $a$, $b$, and $z$ exist \cite{temme83,wimp84}. However, here we deal with
an exceptionally special case of $\mathcal{U}(a,b,z)$ where $b=\frac{1}{2}$, $a>0$ and $z>0$. This suggests that a
dedicated procedure for the calculation of $\mathcal{U}(a,b,z)$ in this special case should be designed. In fact, 
the general algorithm is considerably more complicated and connected with a large computational overhead. Our method 
for the evaluation of $\mathcal{U}(a,1/2,z)$ with $a>0$, $z>0$ is presented in the Supplemental Material \cite{supplement}.
To utilise the 
expression (\ref{eq:suabz}) it is also necessary to compute the Euler Gamma function, $\Gamma(z)$, for the positive 
real values of $z$ but efficient methods exist \cite{stegun72}.

\subsection{Calculation of $S(\alpha,\beta,\gamma)$ for $\alpha \leq -1$}
\label{subsec:silanint}

A slight inconvenience connected with the adopted regularisation scheme for the integrals $S(\alpha,\beta,\gamma)$ is 
that they are not continuous as functions of $\alpha$. Strictly speaking, there are discontinuities for each negative 
integer value of $\alpha$. As a result, negative integer and noninteger values of $\alpha$ have to be treated 
separately which is pursued in the next subsections. Let us note, however, that once the summation such as in Eq.
(\ref{eq:hj00eps}) is performed, the result is a continuous function of all parameters, as it should be.

\subsubsection{Noninteger values of $\alpha$}

By integrating Eq. (\ref{eq:sila}) by parts and dropping the singular terms one obtains the following
recursion relation
\begin{align}
\label{eq:Srec1}
 S(\alpha-1,\beta,\gamma)=-\frac{\beta}{\alpha} S(\alpha,\beta,\gamma) + \frac{2 \gamma}{\alpha} 
 S(\alpha+1,\beta,\gamma),
\end{align}
when $\alpha$ is a noninteger number, $\alpha < -1$. At this point it becomes advantageous to introduce a slightly
simplified family of integrals
\begin{align}
\label{eq:ts}
 T_s^{\pm}(x) = \mathfrak{R} \int_0^\infty \frac{dz}{z^s}\,e^{\pm z - x z^2},
\end{align}
with $x>0$ and (noninteger) real $s>-1$. The above integrals are related to the initial $S(\alpha,\beta,\gamma)$ by an
elementary exchange of variables, giving
\begin{align}
\label{eq:stot}
S(\alpha,\beta,\gamma) = |\beta|^{-\alpha-1} T_{-\alpha}^{\,\rm{sgn}\,\beta} (\gamma/\beta^2),
\end{align}
where $\mbox{sgn}$ denotes the sign function. Starting with Eq. (\ref{eq:Srec1}) it becomes simple to show that the
integrals $T_s^{\pm}(x)$ obey the following recursion relation
\begin{align}
\label{eq:tsrec}
 s\,T_{s+1}^{\pm}(x) = \pm\,T_s^{\pm}(x) - 2x\,T_{s-1}^{\pm}(x),
\end{align}
which is more straightforward to handle than the initial Eq. (\ref{eq:Srec1}). Let us now consider
numerical evaluation of $T_s^{\pm}(x)$ with the help of the recursive process (\ref{eq:tsrec}). Firstly, one requires 
two starting values, $T_{s_0}^{\pm}(x)$ and $T_{s_0-1}^{\pm}(x)$, with $s_0 \in (0,1)$. These integrals are regular and 
can be evaluated with methods described in the previous section. Namely, one has
\begin{align}
\label{eq:tstart}
T_{s_0}^{\pm}(x) = (2\sqrt{x})^{s_0-1} \Gamma(1-s_0)\,\mathcal{U}\left(\frac{1-s_0}{2},\frac{1}{2},\frac{1}{4x}\right),
\end{align}
and similarly for the second starting value. The next issue that needs to be addressed is numerical stability of the 
recursive process (\ref{eq:tsrec}). We tested the reliability of Eq. (\ref{eq:tsrec}) for the calculation of 
$T_s^{\pm}(x)$ up to $s=30$ which is sufficient for all practical purposes, and for a broad range of $x$. The general 
conclusion, both for $T_s^+$ and $T_s^-$, is that Eq. (\ref{eq:tsrec}) is numerically satisfactory only when $x>1$ and 
a significant loss of digits is observed for the smaller values of $x$. However, it is natural to expect that for $x<1$ 
the downward recursion is stable, so let us rewrite Eq. (\ref{eq:tsrec}) as follows:
\begin{align}
\label{eq:rss}
 r_{s-1}^\pm(x) = \frac{2x}{\pm1-s\,r_s^\pm(x)},
\end{align}
where $r_s^\pm(x)=T_{s+1}^\pm(x)/T_s^\pm(x)$ is the Miller-like quotient \cite{gautschi67}. Since $r_s^\pm$ vanishes
quickly with 
increasing $s$ one can initiate the downward recursion at some large $s$ with $r_s^\pm=0$. The recursion is carried out 
until a value, $s_0$, is reached for which $0<s_0<1$. The desired quantities $T_s^\pm$ are then recovered consecutively 
by using the definition of $r_s^\pm$ \emph{i.e.} $T_{s_0+1}^\pm = r_{s_0}^\pm T_{s_0}^\pm$, $T_{s_0+2}^\pm = 
r_{s_0+1}^\pm T_{s_0+1}^\pm$ \emph{etc.}, where the starting value, $T_{s_0}^\pm$, is given by Eq. (\ref{eq:tstart}).

Let us also note that for $\beta$ equal to zero, $S(\alpha,0,\gamma)$, the strategy adopted in 
this subsection is no longer valid since the relation Eq. (\ref{eq:stot}) becomes ill-defined. However, one can return 
to Eq. (\ref{eq:Srec1}) which immediately leads to
\begin{align}
S(\alpha,0,\gamma) = \frac{1}{2} \frac{\Gamma(\frac{a+1}{2})}{\gamma^{\frac{a+1}{2}}}.
\end{align}

\subsubsection{Integer values of $\alpha$}

Similarly as for the noninteger values of $\alpha$, one starts with the definition, Eq. (\ref{eq:sila}), and integrates 
it by parts. However, as now the values of $\alpha$ are restricted to the negative integers, boundary terms in the 
integration by parts give $\varepsilon$-independent, finite contributions. Because of that, the corresponding 
recursion relation is inhomogeneous
\begin{align}
 S(\alpha-1,\beta,\gamma)=
\sum_{l=0}^{-\alpha/2} \frac{\beta^{-\alpha-2l}}{(-\alpha-2l)!}\frac{(-\gamma)^l}{l!}
-\frac{\beta}{\alpha} S(\alpha,\beta,\gamma) + \frac{2 \gamma}{\alpha} 
 S(\alpha+1,\beta,\gamma).
\end{align}
With formally the same definition of $T_s^{\pm}(x)$, Eq. (\ref{eq:ts}), one can derive the corresponding recursion
\begin{align}
\label{eq:tsrec2}
 s\,T_{s+1}^{\pm}(x) =\pm \sum_{l=0}^{s/2}\frac{(-x)^l}{l!(s-2l)!} \pm\,T_s^{\pm}(x) - 2x\,T_{s-1}^{\pm}(x),
\end{align}
which is more compact and numerically convenient. To initiate the recursive process one needs two starting values,
$T_1^{\pm}(x)$ and $T_0^{\pm}(x)$. The expression for the latter is rather simple
\begin{align}
\label{eq:t0}
T_0^{\pm}(x) = \sqrt{\frac{\pi}{4x}}\,e^{\frac{1}{4x}}\left[ 1 \pm \mbox{Erf}\left( \frac{1}{\sqrt{2}x} \right) \right],
\end{align}
where $\mbox{Erf}(*)$ is the error function. To bring the former integral into a more convenient form let us introduce
the following family of functions
\begin{align}
\omega_k(x) = \int_0^\infty dz\, z^k \log(z) e^{-x z - z^2}.
\end{align}
With help of these integrals $T_1^{\pm}(x)$ can be rewritten as
\begin{align}
T_1^{\pm}(x) = \mp\,\omega_0\left(\pm\frac{1}{\sqrt{x}}\right) + 2\,\omega_1\left(\pm\frac{1}{\sqrt{x}}\right).
\end{align}
Note that in contrast with the left-hand-side of the above expression, the integrals present on the right-hand-side
are regular and can be understood in the ordinary sense. We developed a dedicated procedure for the calculation of the
integrals $\omega_0(x)$ and $\omega_1(x)$, which is based on the series expansions and polynomial interpolation. All
necessary expressions are given in the Supplemental Material \cite{supplement}, along with comments and a sketch of the
algorithm.

The remaining issue is the numerical stability of the recursive process (\ref{eq:tsrec2}). For $x>1$ the expression 
(\ref{eq:tsrec2}) can be used as it stands 
\emph{i.e.} in the direction of increasing $s$. However, for $x<1$ it is necessary to apply the downward recursion. 
Unfortunately, since the recursion (\ref{eq:tsrec2}) is inhomogeneous, a trick similar as in Eq. (\ref{eq:rss}) cannot 
be utilised here. Therefore, one has to start at some large $s_{max}$ (significantly larger than the
maximal desired $s$) with $T_{s_{max}+1}^{\pm}(x)=0$ and $T_{s_{max}}^{\pm}(x)=1$, and carry out the recursion downward
(\ref{eq:tsrec2}) until $s=0$ is reached. A slight inconvenience connected with this approach is that the results of
the recursion are not directly $T_s^{\pm}(x)$, but rather $T_s^{\pm}(x)$ multiplied by a constant (the same for each
$s$). To determine this constant one simply evaluates $T_0^{\pm}(x)$ independently by using Eq. (\ref{eq:t0}), compares
with the corresponding result of the downward recursion and rescales all values properly.

\section{Conclusions}
\label{sec:conclusion}
We have presented how six types of two-electron integrals, necessary for the explicitly 
correlated electronic structure calculations, can be reduced to simple quantities. We have applied the 
McMurchie-Davidson scheme first, and reduced all two-electron integrals to the combinations of the derivatives of the basic 
integrals. The differentiations were carried out without specifying the mathematical form of the correlation factor
with the help of the 
Hobson theorem. Finally, the resulting radial derivatives were evaluated recursively starting with a class of
ordinary one-dimensional integrals. 

When the developed scheme is applied to the range-separated correlation factor, all basic quantities can be written 
down as linear combinations of the integrals $S(\alpha,\beta,\gamma)$. Their calculation is not trivial and 
we have discussed in details all necessary numerical procedures that should be applied for specific values of the parameters. We 
would like to stress that virtually all novel expressions presented in this work were extensively checked with the help 
of the symbolic environment of \textsc{Mathematica} package \cite{mathematica07}. Implementation of the production code 
is in progress.

The present work opens up a possibility for the practical use of the range-separated factor in explicitly correlated 
calculations for many-electron atoms and molecules. We believe that the introduction of this new correlation factor 
will noticeably improve the level of accuracy which can routinely be reached with these methods. Extensive benchmarks 
will be reported shortly.

\begin{acknowledgments}
This work was supported by the Polish Ministry of Science and Higher Education, grant NN204 182840.
RM acknowledges the Foundation for Polish Science for the support through the \textit{``Mistrz''} program. We would 
like to thank Bogumi\l~Jeziorski for discussions, as well as reading and commenting on the manuscript.
\end{acknowledgments}


\end{document}


\preprint{AIP/JCP/}

\title[]{\centering{Calculation of the molecular integrals with the range-separated correlation factor. \\Supplemental
material.}}

\author{Micha\l~Silkowski}
\author{Micha\l~Lesiuk}
\email{lesiuk@tiger.chem.uw.edu.pl.}
\author{Robert Moszynski}
\affiliation{Faculty of Chemistry, University of Warsaw, Pasteura 1, 02-093 Warsaw, Poland}

\date{\today}

\begin{abstract}
The present document serves as a supplemental material for the publication \emph{Calculation of the molecular integrals
with the range-separated correlation factor}. It contains data or derivations which are of some importance but are not
necessary for an overall understanding of the manuscript. However, the material presented here will be useful for a
reader who wishes to repeat the derivations in details or implement the formulae independently.
\end{abstract}

\maketitle

\section*{}

\subsection{Transformation between spherical and Cartesian differentials}
\label{app:appa}

An arbitrary mixed Cartesian differential can be rewritten as an adequate linear combination of the spherical tensor 
gradient operators by means of the following formula
\begin{align}
\left(\px\right)^{l_x}\left(\py\right)^{l_y}\left(\pz\right)^{l_z}= N(l,l_x,l_y,l_z) \sum_{l\leq 
l_x+l_y+l_z} c(l,m,l_x,l_y,l_z) \mathcal{Y}_{lm}(\nablab),
\end{align}
where 
\begin{align}
\label{eq:N}
N(l,l_x,l_y,l_z)= \left(  \frac{2^3 \pi (l+1)! (2l_x)!(2l_y)!(2l_z)!}{(2l+2)!l_x!l_y!l_z!} \right)^{\frac{1}{2}},
\end{align}
and
\begin{align}
\begin{split}
\label{eq:cinv}
   c(l,m,l_x,l_y,l_z)=&\sum_{l_{x'}+l_{y'}+l_{z'}=l} 
\frac{(l_x+l_{x'})!(l_y+l_{y'})!(l_z+l_{z'})!}{\left(\frac{l_x+l_{x'}}{2}\right)! \left(\frac{l_y+l_{y'}}{2}\right)! 
\left(\frac{l_z+l_{z'}}{2}\right)!} \\
   &\sqrt{\frac{l! l_x! l_y! l_z! (l-\left| m\right| )!}{(2 l)! (2 l_{x'})! (2 l_{y'})! (2 l_{z'})! (l + \left| 
m\right|)!}} \\
  &\frac{1}{2^l l!} \sum_{i=0}^{(l-\left|m\right|)/2}  \binom{l}{i}  \binom{i}{\frac{l_{x'}+l_{y'}-\left|m\right|}{2}} 
  \frac{(-1)^i (2 l-2 i)!}{(l-\left|m\right|-2i)!} \\
    &\sum_{k=0}^{\frac{l_{x'} + l_{y'} - \left|m\right|}{2}} \binom{\frac{l_{x'}+l_{y'}-\left|m\right|}{2}}{k}  
\binom{\left|m\right|}{l_{x'}-2k} 
    {(-1)}^{ - \frac{(\left| m \right| - l_{x'} + 2k)}{2} }.  
\end{split}
\end{align}
Note, that the term under the outer sum is zero unless $\frac{l_x+l_{x'}}{2}$, $\frac{l_y+l_{y'}}{2}$, and
$\frac{l_z+l_{z'}}{2}$ are the all integers. The above expressions can be compared with the relevant formulae from the 
work of Schlegel \cite{schlegel82}.

\subsection{Computation of $\mathcal{U}(a,1/2,z)$ with $a>0$ and $z>0$}
\label{app:appb}

Before presenting the detailed algorithm let us recall several useful formulae which are obeyed by the desired 
quantities, $\mathcal{U}(a,1/2,z)$. Details and comments on these expressions can be found in Ref. \onlinecite{stegun72}
and references therein. Firstly, the small $z$ expansion of the Kummer's hypergeometric function, $M(a,b,z)$, is
\begin{align}
\label{eq:msmallz}
M(a,b,z) = \sum_{n=0}^\infty \frac{(a)_n z^n}{(b)_n n!},
\end{align}
where $(x)_n$ are the Pochhammer symbols. The corresponding expansion for the Tricomi's hypergeometric function can be 
written in terms of $M(a,b,z)$ as
\begin{align}
\label{eq:usmallz}
\mathcal{U}(a,1/2,z) = \sqrt{\pi} \bigg[ \frac{M(a,1/2,z)}{\Gamma(a+1/2)} 
-2\sqrt{z}\,\frac{M(a+1/2,3/2,z)}{\Gamma(a)} \bigg].
\end{align}
Let us also recall the following recursive formulae for increasing the value of $a$
\begin{align}
\label{eq:urec}
 \mathcal{U}(a-1,1/2,z)+(1/2-2a-z)\,\mathcal{U}(a,1/2,z)+a(a+1/2)\,\mathcal{U}(a+1,1/2,z)=0,
\end{align}
and the corresponding continued fraction (CF)
\begin{align}
\begin{split}
\label{eq:ucf}
&a\,\frac{\mathcal{U}(a+1,1/2,z)}{\mathcal{U}(a,1/2,z)} = \frac{a_1}{b_1+}\,\frac{a_2}{b_2+}\cdots,\\
&a_n = -(a+n-1/2)(a+n-1),\\
&b_n = (1/2-2a-z-2n).
\end{split}
\end{align}
Note additionally, that for very large values of $z$ the Poincar\'{e}-type asymptotic formula can be used to simplify
the computation. This formula is a simple inverse powers series in $z$
\begin{align}
\label{eq:uasym}
\mathcal{U}(a,1/2,z) = z^{-a} \sum_{s=0}^\infty \frac{(a)_s(a+1/2)_s}{s!}\frac{1}{(-z)^s},
\end{align}
and can be efficiently implemented.

Let us now present the complete algorithm for calculation of $\mathcal{U}(a,1/2,z)$. First of all, when $z<1/2$ the 
small $z$ expansion is supposed to be used, see Eqs. (\ref{eq:msmallz}) and (\ref{eq:usmallz}). Typically, these series 
converge fast in this regime and give accurate results despite a cancellation between two terms in 
Eq. (\ref{eq:usmallz}). An exception is the region when the value of $a$ is very large but such case is unlikely 
to occur in our applications.

The second ingredient of the algorithm is supposed to be used when $z>1/2$. Let us assume we would like to compute 
$\mathcal{U}(a,1/2,z)$ for a fixed $z$, up to some predefined value $a_{max}$. Firstly, we compute the ratio 
$\mathcal{U}(a_{max},1/2,z)/\mathcal{U}(a_{max}-1,1/2,z)$ by using the continued fraction formula (\ref{eq:ucf}). 
The evaluation can be performed with the Steed's algorithm and convergence of CF is acceptable for virtually every
$z>1/2$ and any reasonable 
$a_{max}$. The only inconvenience is the fact that the consecutive numerators and denominators in Eq. (\ref{eq:ucf}) 
grow very fast with $n$ and need to be scaled by a small number from time to time to avoid overflows. Once the ratio 
$\mathcal{U}(a_{max},1/2,z)/\mathcal{U}(a_{max}-1,1/2,z)$ has been obtained, we rewrite the recursion relation 
(\ref{eq:urec}) as follows
\begin{align}
\label{eq:urec2}
 r_{a-1}(z)=-\frac{1}{(1/2-2a-z)+a(a+1/2)r_a(z)},
\end{align}
where the ratio $r_a(z) = \mathcal{U}(a+1,1/2,z)/\mathcal{U}(a,1/2,z)$. One sees that the value obtained from the CF 
formula is simply 
$a_{max}\,r_{a_{max}-1}$. Starting with this value we carry out the recursion (\ref{eq:urec2}) downwards, until a 
minimal value, $a_0$, is reached which satisfies $0<a_0<1$. At this point, $\mathcal{U}(a_0,1/2,z)$ can be computed
with expressions Eqs. (\ref{eq:msmallz}) and (\ref{eq:usmallz}) for $z<10$ and Eq. (\ref{eq:uasym}) for $z>10$ since
the corresponding series converge acceptably for small $a$. Finally, using the definition of
$r_{a_0}$ one can see that $\mathcal{U}(a_0+1,1/2,z)=r_{a_0}\mathcal{U}(a_0,1/2,z)$, then 
$\mathcal{U}(a_0+2,1/2,z)=r_{a_0+1}\mathcal{U}(a_0+1,1/2,z)$ and so forth, until the 
value of $a_{max}$ is reached. This furnishes the computation of the $\mathcal{U}(a,1/2,z)$ functions. 

\subsection{Basic integrals with the range-separated correlation factor}
\label{app:appc}
Before expressing all basic integrals, $B_i$, trough the $S(\alpha,\beta,\gamma)$ integrals, let us introduce a
notation that makes the expressions more clear and succinct. Supposing one has two distinct correlation factors
in the form given by Eq. (54) in the paper, with two sets of the parameters: $c_0,\rho,B,\mu$ and $c_0',\rho',B',\mu'$,
respectively. Firstly, the operator $\wp'$ acting on an expression changes all unprimed quantities into their primed
counterparts (and \emph{vice-versa}), for instance:
\begin{align}
 f_{12}'(r) =\wp' f_{12}(r) = \left(1+\frac{1}{2}r\right)e^{-\mu' r^2} + c_0' \left(1-e^{-\mu' r^2}\sum_{\kappa=0}^n
\frac{{\mu'}^{\kappa} r^{2 \kappa}}{\kappa!}\right)r^{\rho'} e^{B' r}.
\end{align}
Hereafter, the notation $\Big[ \ldots \Big]_{-}$ (and $\Big\{\ldots \Big\}_{-}$), should be considered as exactly the
same expression as in the bracket before, but with $\xi\rightarrow -\xi$ in the second argument of S \emph{e.g.}
\begin{align}
 S(\rho,B+2\xi Q,\xi)+\bigsq =  S(\rho,B+2\xi Q,\xi) + S(\rho,B-2\xi Q,\xi).
\end{align}
Finally, $S(\alpha,...)$ should be considered as integral $S$ with exactly the same
$\beta$ and $\gamma$ arguments as in the preceding integral $S$, for example:
\begin{align}
 S(\alpha,\beta,\gamma)+S(\alpha',...)+S(\alpha'',...)=
S(\alpha,\beta,\gamma)+S(\alpha',\beta,\gamma)+S(\alpha'',\beta,\gamma).
\end{align}
With help of this notation one can represent the basic integrals as follows (note that $B_5$ expresses similarly as
$B_3$ and thus a separate treatment is not required here):
\newpage
\begin{align}
\begin{split}
B_1=\sqrt{\frac{\pi^5}{p+q}}\frac{1}{qp} \frac{1}{Q}e^{-\xi Q^2} \Big[S(0,2\xi Q,\xi)-S(0,-2\xi
Q,\xi)\Big].
\end{split}
\end{align}

\begin{align}
\begin{split}
\label{eq:I2byS}
 B_2 &= \sqrt{\frac{\pi^5}{p+q}}\frac{1}{qp} \frac{1}{Q}e^{-\xi Q^2} \Bigg[ S(0,2\xi Q,\xi+\mu) +\frac{1}{2} S(1,2\xi
Q,\xi+\mu) + c_0 S(\rho,B+2\xi Q,\xi)
 \Bigg.
 \\ &-\Bigg. c_0\sum_{\kappa=0}^{n} \frac{\mu^{\kappa}}{\kappa !} S(\rho+2\kappa,B+2\xi Q,\xi+\mu) \Bigg] + \Bigg[
\ldots \Bigg]_{-},
 \end{split}
\end{align}

\begin{align}
\begin{split}
\label{eq:I3byS}
  B_3 &= \sqrt{\frac{\pi^5}{p+q}}\frac{1}{qp} \frac{1}{Q}e^{-\xi Q^2} \Bigg[ S(1,2\xi Q,\xi+\mu) + \frac{1}{2} S(2,2\xi
Q,\xi+\mu) + c_0 S(\rho+1,B+2\xi Q,\xi) \Bigg.
 \\ &-\Bigg. c_0\sum_{\kappa=0}^{n} \frac{\mu^{\kappa}}{\kappa !} S(\rho+2\kappa+1,B+2\xi Q,\xi+\mu) \Bigg]
+  \Bigg[ \ldots \Bigg]_{-},
 \end{split}
\end{align}

\begin{align}
\begin{split}
 B_4&= \sqrt{\frac{\pi^5}{p+q}}\frac{1}{qp} \frac{1}{Q}e^{-\xi Q^2} \Bigg\{ S(1,2\xi Q,\xi +\mu + \mu') +
\left(1+\wp'\right)\frac{1}{2}S(2,...) + \frac{1}{4}S(3,...) +
 \\&+ \left(1+ \wp' \right) c_0 \left[ S(\rho+1,2\xi Q+B,\xi+\mu') +\frac{1}{2} S(\rho+2,...) \right] - \\
 &-\left(1+ \wp' \right) c_0 \sum_{\kappa=0}^n \frac{\mu^{\kappa}}{\kappa!} \left[ S(\rho+2\kappa+1,2\xi
Q+B,\xi+\mu'+\mu) +\frac{1}{2} S(\rho+2\kappa+2,...) \right] +\\
 &+ c_0 c_0' \sum_{k=0}^n \sum_{k'=0}^{n'} \frac{\mu^{k}{\mu'}^{k'}}{k!{k'}!} S(\rho+\rho'+2k+2k'+1,2\xi Q +B+ B',\xi
+\mu+\mu') + \\
 &+ c_0 c_0' S(\rho+\rho'+1,2\xi Q+B+B',\xi) \Bigg\} + \Bigg\{ \ldots \Bigg\}_{-},
\end{split}
\end{align}

\begin{align}
\begin{split}
 B_6 &= \sqrt{\frac{\pi^5}{p+q}}\frac{1}{qp} \frac{1}{Q}e^{-\xi Q^2} \Bigg\{ 
 \frac{1}{4}S(1,2\xi Q, \xi + \mu + \mu') + 4\mu \mu' S(3,...) + \mu \mu' S(5,...) + 4 \mu \mu' S(4,...)+ \\
 &+ (1+ \wp')\Big( -\mu  S(2,...)- \frac{\mu}{2} S(3,...)\Big) + 
 (1+\wp')\bigg[c_0'  \Big( \frac{1}{2} B' S(\rho'+1,2\xi Q + B',\xi + \mu) + \\
 &+ \frac{1}{2}\rho' S(\rho',...) - 2\mu B' S(\rho'+2,...) - \mu B' S(\rho'+3,...) - 2\mu \rho' S(\rho'+1,...) - \mu
\rho' S(\rho'+2,...) \Big) \bigg] + \\ 
 &+ c_0 c_0' \bigg[ B B' S(\rho + \rho' +1,2 \xi Q + B + B',\xi) + \rho \rho' S(\rho + \rho' -1,...) + \\
 &+ (1+\wp') \Big( B \rho' S(\rho + \rho',...) \Big) \bigg] + 
 (1+\wp') \bigg[ c_0 \sum_{\kappa=0}^n \frac{\mu^\kappa}{\kappa!} \\
 & \Big( \frac{B}{2} S(\rho+2\kappa+1,2\xi Q + B,\xi + \mu + \mu') -\mu S(\rho+2\kappa+2,...) + \kappa
S(\rho+2\kappa,...) - \\
 &- 2\mu' B S(\rho+2\kappa+2,...) + 2 \mu \mu' S(\rho+2\kappa+3,...) -4\mu' \kappa S(\rho+2\kappa+1,...) - \\
 &- \mu' B S(\rho+2\kappa+3,...) +2\mu \mu' S(\rho+2\kappa+4,...) -\mu' \kappa S(\rho+2\kappa+2,...) \Big) \bigg]  + \\
 &+ c_0 c_0' (1+\wp') \bigg[ \sum_{\kappa=0}^n \frac{{\mu}^\kappa}{\kappa!} \Big( B B' S(\rho+\rho'+2\kappa +1,2 \xi Q +
B + B',\xi + \mu) + \\
 &+ 2 \mu B' S(\rho+\rho'+2\kappa +2,...) + 2 \kappa B' S(\rho+\rho'+2\kappa,...) + \rho' B S(\rho+\rho'+2\kappa,...) -
\\
 &- 2 \mu \rho' S(\rho+\rho'+2\kappa+1,...) + 2 \kappa \rho' S(\rho+\rho'+2\kappa-1,...) \Big)  \bigg] + \\
 &+ c_0 c_0' \sum_{\kappa=0}^n \sum_{\kappa'=0}^{n'} \frac{{\mu}^\kappa {\mu'}^{\kappa'}}{\kappa! {\kappa'}!} \bigg[ B
B' S(\rho+\rho'+2(\kappa+\kappa')+1,2 \xi Q + B + B',\xi +\mu + \mu') + \\
 &+ 4 \mu \mu' S(\rho+\rho'+2(\kappa+\kappa')+3,...) + 4 \kappa \kappa' S(\rho+\rho'+2(\kappa+\kappa')-1,...) + \\
 &+ (1+\wp')\Big( -2\mu B' S(\rho+\rho'+2(\kappa+\kappa')+2,...) -4\mu \kappa' S(\rho+\rho'+2(\kappa+\kappa')+1,...) +
\\
 &+ 2 B' \kappa S(\rho+\rho'+2(\kappa+\kappa'),...) \Big) \bigg] \Bigg\} + \Bigg\{ \ldots \Bigg\}_{-},
\end{split}
\end{align}

\subsection{Calculation of the $\omega_k$ integrals}
\label{app:appd}

Before passing to the actual algorithm let us note that for evaluation of the desired molecular integrals one requires
$\omega_0(x)$ and $\omega_1(x)$ with arbitrary real $x$. Herein, we present the algorithm for $x>0$ only since
the treatment of negative values of $x$ is completely analogous. Considering the evaluation of $\omega_0(x)$ and
$\omega_1(x)$ let us first note that the following small $x$ expansions hold
\begin{align}
\omega_0(x) = \sum_{k=0}^\infty \frac{(-x)^k}{k!}
\Gamma\left(\frac{k+1}{2}\right)\psi\left(\frac{k+1}{2}\right),\\
\omega_1(x) = \sum_{k=0}^\infty \frac{(-x)^k}{k!} \Gamma\left(\frac{k}{2}+1\right)\psi\left(\frac{k}{2}+1\right),
\end{align}
where $\psi(z)$ is the digamma function. The above expressions can be derived by simply expanding the
exponential term in the integrand and integrating term by term. These series converge reasonably well for small $x$ and 
the coefficients can be tabulated in advance or computed on the fly recursively. Let us also recall that for the 
half-integer values the digamma function takes the following closed form
\begin{align}
 \psi(n+1/2) = -\gamma_E-2\log 2+\sum_{k=1}^n \frac{2}{2k-1},
\end{align}
where $\gamma_E$ is the Euler-Mascheroni constant.

By performing a linear exchange of variables the integral $\omega_0(x)$ can be rewritten as
\begin{align}
\omega_0(x) = \int_0^\infty dz\, \log(z) e^{-x z - z^2} = \frac{1}{x}\int_0^\infty dt\,\log t\,
e^{-t-t^2/x^2}-\frac{\log x}{x}\int_0^\infty dt\,e^{-t-t^2/x^2}.
\end{align}
By expanding the exponential terms containing $1/x^2$ in both integrals the following large $x$ expansion is obtained
\begin{align}
\label{eq:omega0asym}
\omega_0(x) = \sum_{k=0}^\infty \frac{(-1)^k}{x^{2k+1}} \frac{(2k)!\psi(2k+1)}{k!}-\log x \sum_{k=0}^\infty
\frac{(-1)^k}{x^{2k+1}} \frac{(2k)!}{k!}.
\end{align}
The above series is clearly divergent \emph{i.e.} it provides asymptotic information about $\omega_0(x)$ when $x$ is
large. In practice, the above expression is very useful for calculations with $x>50$, say, when typically only several 
terms are necessary to converge to the machine precision. Calculation of the individual terms is fairly straightforward,
taking into account that for the integer values of $n$ the digamma function is given by
\begin{align}
\psi(n) = \sum_{k=1}^{n-1} \frac{1}{k}-\gamma_E.
\end{align}
A completely analogous expression to Eq. (\ref{eq:omega0asym}) holds for the second integral, $\omega_1(x)$.

Having resolved the cases of small and large $x$, let us discuss the calculation of $\omega_0(x)$ and $\omega_1(x)$ for
moderate $x$ where neither of the above schemes give a sufficient accuracy. After numerical experiments we found that 
the following expansions give reasonably accurate piecewise fits for the desired quantities
\begin{align}
\label{eq:ffit}
 F_n(x) = \sum_{k=0}^n \frac{c_k}{x^k} + \log x \sum_{k=0}^n \frac{d_k}{x^k}.
\end{align}
Our scheme for calculation of $\omega_0(x)$ and $\omega_1(x)$ goes as follows: for $x<3$ we use the small-$x$ infinite
series expansion described above, where it converges to the accuracy of 15 digits in less than 50 terms. Obviously, for
$x>50$ the aforementioned large-$x$ asymptotic expansion is used, which gives the machine precision after about
20 terms. For the remaining values of $x$ we use the following approximations: 
\begin{itemize}
 \item $3<x<5$: $F_n(x)$ with $n=6,5$ for $\omega_0(x)$ and $\omega_1(x)$, respectively,
 \item $5<x<10$: $F_n(x)$ with $n=6$ for both $\omega_0(x)$ and $\omega_1(x)$,
 \item $10<x<20$: $F_n(x)$ with $n=5$ for both $\omega_0(x)$ and $\omega_1(x)$,
 \item $20<x<50$: $F_n(x)$ with $n=4,5$ for $\omega_0(x)$ and $\omega_1(x)$, respectively.
\end{itemize}
The coefficients $c_k$, $d_k$ are different for each interval and are obtained from the least-squares fit on a equally 
spaced ($\Delta=0.001$) grid. These reference data were obtained by integrating numerically $\omega_0(x)$ and 
$\omega_1(x)$ in an extended arithmetic precision. In the following Tables we report the values of the coefficients 
$c_k$, $d_k$ for each interval. Accuracy of these fits is at least 13 significant digits and much better on the average, 
which is sufficient for the present purposes. Tests were performed on a large sample of randomly generated points.

\begin{ruledtabular}
\begin{table}
\caption{Coefficients $c_k$ and $d_k$ in the analytical fits of $\omega_0(x)$ and $\omega_1(x)$ with the functional 
form given by Eq. (\ref{eq:ffit}) for $3<x<5$. The symbol $[k]$ denotes the powers of 10, 10$^k$.}
\label{table1}
\begin{tabular}{ccc}
 $n$ & $c_n$ & $d_n$ \\
\hline
 & \multicolumn{2}{c}{$\omega_0(x)$} \\
\hline
 0   & $-$2.79 418 516 254 156[$-$01] & $+$4.35 827 381 193 435[$-$02] \\
 1   & $-$2.90 713 893 948 490[$+$01] & $+$6.03 123 542 136 014[$+$00] \\ 
 2   & $-$4.95 348 276 834 514[$+$02] & $+$1.90 103 007 336 098[$+$02] \\ 
 3   & $-$1.87 541 627 996 346[$+$03] & $+$1.37 709 919 986 376[$+$03] \\ 
 4   & $-$2.97 088 892 758 365[$+$02] & $+$2.88 186 650 129 034[$+$03] \\ 
 5   & $+$2.15 515 331 846 835[$+$03] & $+$1.58 883 573 230 859[$+$03] \\ 
 6   & $+$5.41 315 200 857 445[$+$02] & $+$1.45 477 988 235 751[$+$02] \\
\hline
 & \multicolumn{2}{c}{$\omega_1(x)$} \\
\hline
 0   & $-$6.81 608 673 256 649[$-$03] & $-$7.81 053 598 547 109[$-$04] \\ 
 1   & $-$6.03 136 412 874 311[$+$00] & $+$1.26 540 880 331 240[$+$00] \\
 2   & $-$1.23 965 414 329 573[$+$02] & $+$4.59 860 722 344 580[$+$01] \\
 3   & $-$2.92 393 636 456 307[$+$02] & $+$2.90 445 255 218 630[$+$02] \\
 4   & $+$2.42 606 911 139 842[$+$02] & $+$3.44 114 756 369 323[$+$02] \\
 5   & $+$1.79 651 772 296 869[$+$02] & $+$5.59 224 732 011 467[$+$01] \\
\end{tabular} 
\end{table} 
\end{ruledtabular}

\begin{ruledtabular}
\begin{table}
\caption{Coefficients $c_k$ and $d_k$ in the analytical fits of $\omega_0(x)$ and $\omega_1(x)$ with the functional 
form given by Eq. (\ref{eq:ffit}) for $5<x<10$. The symbol $[k]$ denotes the powers of 10, 10$^k$.}
\label{table2}
\begin{tabular}{ccc}
 $n$ & $c_n$ & $d_n$ \\
\hline
 & \multicolumn{2}{c}{$\omega_0(x)$} \\
\hline
 0   & $+$2.67 336 306 337 877[$-$02] & $-$3.83 390 302 078 230[$-$03] \\
 1   & $+$4.47 174 316 838 360[$+$00] & $-$2.09 112 260 860 070[$+$00] \\
 2   & $+$1.69 542 503 617 498[$+$02] & $-$5.29 464 781 748 209[$+$01] \\
 3   & $+$1.39 875 310 206 529[$+$03] & $-$7.03 047 033 787 913[$+$02] \\
 4   & $+$2.02 964 879 213 659[$+$03] & $-$2.77 647 534 682 290[$+$03] \\
 5   & $-$2.17 744 743 464 807[$+$03] & $-$2.80 275 812 247 144[$+$03] \\
 6   & $-$1.42 551 927 395 417[$+$03] & $-$4.46 967 011 675 890[$+$02] \\
\hline
 & \multicolumn{2}{c}{$\omega_1(x)$} \\
\hline
 0   & $+$2.89 074 504 802 456[$-$02] & $-$4.23 814 942 235 530[$-$03] \\
 1   & $+$4.44 902 630 985 788[$+$00] & $-$1.00 783 068 266 863[$+$00] \\
 2   & $+$1.08 226 256 016 102[$+$02] & $-$3.85 434 024 325 027[$+$01] \\
 3   & $+$4.27 985 618 179 230[$+$02] & $-$3.17 198 102 880 598[$+$02] \\
 4   & $-$7.33 719 621 295 158[$+$02] & $-$3.53 983 894 638 774[$+$02] \\
 5   & $-$6.13 015 358 740 435[$+$02] & $+$7.33 534 471 812 601[$+$02] \\
 6   & $+$8.04 702 720 006 004[$+$02] & $+$3.22 960 340 787 190[$+$02] \\
\end{tabular} 
\end{table} 
\end{ruledtabular}

\begin{ruledtabular}
\begin{table}
\caption{Coefficients $c_k$ and $d_k$ in the analytical fits of $\omega_0(x)$ and $\omega_1(x)$ with the functional 
form given by Eq. (\ref{eq:ffit}) for $10<x<20$. The symbol $[k]$ denotes the powers of 10, 10$^k$.}
\label{table3}
\begin{tabular}{ccc}
 $n$ & $c_n$ & $d_n$ \\
\hline
 & \multicolumn{2}{c}{$\omega_0(x)$} \\
\hline
 0   & $+$9.85 170 564 088 696[$-$04] & $-$1.34 352 155 222 340[$-$04] \\
 1   & $-$3.14 495 326 884 370[$-$01] & $-$1.05 330 699 931 180[$+$00] \\
 2   & $+$1.14 410 655 314 381[$+$01] & $-$3.35 296 508 844 390[$+$00] \\
 3   & $+$1.01 789 460 952 360[$+$02] & $-$4.91 298 273 165 823[$+$01] \\
 4   & $+$5.99 368 814 053 674[$+$01] & $-$1.75 836 403 023 178[$+$02] \\
 5   & $-$1.74 101 744 430 719[$+$02] & $-$7.11 820 445 035 004[$+$01] \\
\hline
 & \multicolumn{2}{c}{$\omega_1(x)$} \\
\hline
 0   & $+$1.79 379 586 198 002[$-$04] & $-$2.67 961 445 256 187[$-$05] \\
 1   & $+$4.80 466 657 853 673[$-$03] & $-$2.80 747 000 711 979[$-$03] \\
 2   & $-$2.08 437 134 302 115[$+$00] & $-$5.01 274 816 607 827[$-$01] \\
 3   & $-$6.61 049 670 065 573[$+$01] & $+$2.30 089 102 665 028[$+$01] \\
 4   & $-$2.02 378 315 248 675[$+$02] & $+$1.79 404 122 905 036[$+$02] \\
 5   & $+$2.61 597 803 283 919[$+$02] & $+$1.59 307 953 057 555[$+$02] \\
\end{tabular} 
\end{table} 
\end{ruledtabular}

\begin{ruledtabular}
\begin{table}
\caption{Coefficients $c_k$ and $d_k$ in the analytical fits of $\omega_0(x)$ and $\omega_1(x)$ with the functional 
form given by Eq. (\ref{eq:ffit}) for $20<x<50$. The symbol $[k]$ denotes the powers of 10, 10$^k$.}
\label{table4}
\begin{tabular}{ccc}
 $n$ & $c_n$ & $d_n$ \\
\hline
 & \multicolumn{2}{c}{$\omega_0(x)$} \\
\hline
 0   & $+$1.39 870 024 574 414[$-$05] & $-$1.76 030 960 012 356[$-$06] \\
 1   & $-$5.70 280 882 024 327[$-$01] & $-$1.00 126 191 147 610[$+$00] \\
 2   & $+$5.28 077 445 114 726[$-$01] & $-$1.35 267 549 073 835[$-$01] \\
 3   & $+$5.54 783 065 129 888[$+$00] & $-$1.14 510 519 374 110[$+$00] \\
 4   & $-$2.07 952 101 488 230[$+$00] & $-$1.13 489 580 958 509[$+$01] \\
\hline
 & \multicolumn{2}{c}{$\omega_1(x)$} \\
\hline
 0   & $-$8.04 790 438 983 053[$-$06] & $+$9.92 668 209 243 167[$-$07] \\
 1   & $-$4.83 240 835 649 057[$-$03] & $+$8.49 304 239 905 638[$-$03] \\
 2   & $-$4.74 103 840 956 836[$-$02] & $-$8.86 737 271 752 571[$-$01] \\
 3   & $-$9.57 722 093 234 973[$+$00] & $+$3.53 498 860 651 889[$+$00] \\
 4   & $-$2.32 473 977 612 046[$+$01] & $+$2.77 433 057 711 969[$+$01] \\
 5   & $+$4.72 702 979 536 110[$+$01] & $-$5.06 583 892 061 286[$+$00] \\
\end{tabular} 
\end{table} 
\end{ruledtabular}